\documentclass[aps,pre,reprint]{revtex4-1}
\usepackage{amsmath}
\usepackage{amssymb}
\usepackage{blindtext}
\usepackage{braket}
\usepackage{comment}
\usepackage{graphicx}
\usepackage{hyperref}
\usepackage{mathrsfs}
\usepackage{mathtools}
\usepackage{subcaption}
\usepackage{tikz}
\usepackage{xcolor}

\usetikzlibrary{positioning}
\usetikzlibrary{calc}
\usetikzlibrary{arrows,shapes,backgrounds}

\DeclareCaptionJustification{justified}{\leftskip=0pt \rightskip=0pt \parfillskip=0pt plus 1fil}
\begin{document}
\title{The Hungry Daemon: An energy-harvesting active particle must obey the Second Law of Thermodynamics}
\author{S.~Bienewald$^1$, D.M.~Fieguth$^2$ and J.R.~Anglin$^2$}

\affiliation{$^1$\mbox{Fachbereich Mathematik,} \mbox{Humboldt-Universit\"{a}t zu Berlin,} \mbox{D-10099 Berlin, Germany}
\\
$^2$\mbox{State Research Center OPTIMAS and Fachbereich Physik,} \mbox{Rheinland-Pf\"alzische Technische Universit\"at,} \mbox{D-67663 Kaiserslautern, Germany}}
\date{\today}

\begin{abstract}
    Thought experiments like Maxwell's Demon or the Feynman-Smoluchowski Ratchet can help in pursuing the microscopic origin of the Second Law of Thermodynamics. Here we present a more sophisticated physical system than a ratchet, consisting of a Hamiltonian active particle which can harvest energy from an environment which may be in thermal equilibrium at a single temperature. We show that while a phenomenological description would seem to allow the system to operate as a Perpetual Motion Machine of the Second Kind, a full mechanical analysis confirms that this is impossible, and that perpetual energy harvesting can only occur if the environment has an energetic population inversion similar to a lasing medium.
\end{abstract}

\maketitle

\section{Introduction}

\subsection{Demons, Ratchets, and the Second Law}
The Laws of Thermodynamics are in no doubt, but the relationship of the Second Law to microscopic physics is still not fully clear. The Second Law cannot really be an independent axiom of causality in its own right, since deterministic mechanics already prescribes everything that can ever happen, and never requires any electron to consult thermodynamics in order to learn what to do next.
General proof of the Second Law as a theorem within mechanics remains elusive, however. In fact, it still seems possible that thermodynamics is \emph{not} only due to microscopic equations of motion, but may instead reflect a constraint on the initial conditions of the universe.

Thought experiments can help focus our understanding by posing concrete scenarios in which the Second Law might reasonably seem to be violated. The oldest example is Maxwell's Demon \cite{Maxwell}: Its apparent ability to violate the Second Law disappears if the vague assumption that the Demon ``observes'' molecular speeds is replaced with interaction between the Demon and the molecules via a physical medium which can itself carry entropy \cite{e19060240}. Another proposal for violating the Second Law, the Feynman-Smoluchowski Ratchet \cite{Feynmanlectures,Smoluchowski}, replaces the Demon's conscious observation and control with a simple mechanical pawl; recognizing that the pawl is not just a one-way valve for motion, but a degree of freedom with its own fluctuations, again restores the Second Law. Even when general fluctuation theorems \cite{Crooks1999EntropyPF,Jarzynski1996NonequilibriumEF,Maes2005SecondLO} assure us \emph{that} we will never be able to violate the Second Law by coupling dynamical mechanisms to canonical reservoirs, concrete scenarios like the Demon and the Ratchet provide useful insights by illustrating precisely \emph{how} the laws of mechanics enforce those of Thermodynamics.

\subsection{Active Particles and Perpetual Motion}
In this paper we therefore offer another scheme for potentially violating the Second Law, which naively seems plausible but which we show will not actually work under the laws of Mechanics. Our example is a more complex mechanism than a pawl: it is a concrete mechanical representation of a so-called \emph{active particle} \cite{fieguth2024dynamical,fieguth2022hamiltonian}, which draws upon an internal energy ``depot'' to move against an external force (often friction). Such active particles can be considered as models for mobile micro-organisms, and like micro-organisms which keep moving by finding food in their environment, active particle models can replenish their internal depots by harvesting energy \cite{ebeling1999active,schweitzer1998complex}.

We therefore consider an active particle which is able to replenish its internal energy by harvesting energy from its environment, as micro-organisms do by eating. A phenomenological model of how this ``hungry'' particle could be expected to behave, and which might indeed offer a good phenomenological description of real energy harvesting under some circumstances, is then shown to allow a Perpetual Motion Machine of the Second Kind, violating the Second Law by moving the particle indefinitely against an external force while drawing energy from an environment which is in equilibrium at a single positive temperature. Describing both the active particle itself and its energy harvesting within valid microscopic mechanics, however, we find instead that the particle can only keep doing work perpetually if its environmental ``food'' has a \emph{negative} temperature. 

This highlights the importance of the fact that environments from which real active particles do harvest energy are \emph{not} in thermal equilibrium, but hold scattered concentrations of energy that are too common to be thermal fluctuations. We also see from this example that energy harvesting is fundamentally just as constrained by thermodynamics as work output is, because both work and harvesting are fundamentally constrained by microscopic mechanics.

\section{Energy harvesting and the Second Law}
\label{sec:phenomenological_approach}
\subsection{Energy from ``crumbs''}
An active particle can move against an opposing force, whether systematic or frictional, by using energy stored in the particle's internal energy depot. Since the depot is finite, and moving against an external force requires power, the active particle must eventually deplete its depot and stop moving actively---unless the active particle can replenish its depot by harvesting energy from its environment. The active particle's motion may enable depot refill by bringing it into contact with new localized energy sources that it can use. We will refer to such localized energy sources, distributed at different locations throughout the environment, as ``crumbs''. If an active particle can find and exploit enough crumbs fast enough, compared to the power output it needs to keep moving, then it should be able to keep moving indefinitely.

If we take the crumbs a bit more seriously as physical objects, rather than just as a way of saying that the particle keeps going, then we have to consider them having their own dynamical degrees of freedom, which can possess a range of possible energies. If the active particle takes energy from a crumb, the crumb loses the same amount of energy. 

It is certainly possible, and could well be realistic, for the initial energies of crumbs to vary thermally, in a Boltzmann distribution. If the crumbs are too cold, their generally low energy levels might be an inadequate food supply for the active particle. As long as the crumbs have a high enough positive temperature, however, they must on average be a rich source of energy, even if some of them are still energy-poor. As long as the crumbs are common enough in the environment, then, it seems plausible that the active particle could continue drawing enough energy from energy-rich crumbs to keep going indefinitely, regardless of the fact that the crumb energies have a Boltzmann distribution. The crumb temperature simply has to be high enough.

\subsection{Perpetual Motion}
This seemingly realistic scenario is in fact impossible: It would violate the Second Law of Thermodynamics, by allowing the operation of a Perpetual Motion Machine of the Second Kind, which takes heat from a reservoir at a single temperature and converts it completely into work.

It does not matter that the crumbs are not supposed to be a thermal reservoir themselves, or that the energy which the active particle takes from them is not supposed to be heat. It does not matter that the active particle and its crumbs are all supposed to be microscopic, or that the active particle is considered to be operating far from equilibrium. The Second Law not only forbids situations which violate it explicitly: it also forbids any scenario which could potentially be exploited to violate the Second Law explicitly. 

If the crumb energies are thermally distributed, then, after being drained by the active particle, they could be raised back again to their initial thermal distribution just by putting them into thermal contact with a heat reservoir at the appropriate temperature. The active particle could then replenish itself from the same crumbs again, and we could repeat this cycle indefinitely, to realize a Perpetual Motion Machine of the Second Kind. The crumbs with thermally distributed initial energies may not themselves be the reservoir in this Machine, but if they can replenish an active particle, then they would enable the Machine to work. So, it must not be possible to harvest energy from crumbs with thermally distributed energies.

\subsection{A paradox?}
This seems at least somewhat surprising. Real active particles do exist, after all---micro-organisms, for example---which really do sustain active motion over long times by consuming food. As a thermodynamical thought experiment, however, an energy-harvesting active particle is indeed somewhat like Maxwell's Demon, in that its ability to harvest energy is described phenomenologically rather than microscopically, like the Demon's ability to see molecules. By providing a microscopic model for the energy-harvesting active particle, we will resolve the conflict between thermodynamics and intuition, and learn a bit more about energy harvesting as well as about the relationship between thermodynamics and mechanics.

\section{Hamiltonian Daemon}
\label{sec:hamiltonian_daemon}
Our mechanical model for energy harvesting by an active particle is based on a previous mechanical model for an active particle. The \textit{Hamiltonian Daemon} is a mechanical model introduced in \cite{PhysRevE.94.042127} which is able to transfer energy from fast to slow degrees of freedom, using a nonlinear resonance. The model is referred to as a ``daemon'' in analogy to the autonomous background processes in Unix-family computer operating systems. In \cite{fieguth2024dynamical} it was shown how this system has the properties of an active particle: It can sustain motion against external forces by drawing energy from an internal depot. In \cite{fieguth2022hamiltonian} it was further shown that this mechanical model can still operate as an active particle under strong damping, with Eulerian motion. 

To focus on the problem of energy harvesting, we will consider here the undamped Daemon, and add to it interactions with crumbs that will allow this ``hungry daemon'' to ``feed''. We first review the ``Daemon'' model itself.

\subsection{The Daemon Hamiltonian}
The Hamiltonian of the Daemon model of \cite{PhysRevE.94.042127} is a particular case of the general form
\begin{align}
    H_\mathrm{D} = H_{\mathrm{M}} + H_{\mathrm{I}} + \gamma H_{\mathrm{C}}, \label{hamiltonian_daemon}
\end{align}
where $H_{\mathrm{M}}$ is the motional energy (kinetic and potential) of the particle, $H_{\mathrm{I}}$ is the particle's internal energy depot and $H_{\mathrm{C}}$ describes the coupling between the previous two subsystems. The parameter $\gamma$ is in general small, so that a potentially large amount of ``fuel'' energy can be converted gradually, over a long time, into a large amount of work in moving the particle against an opposing force. Making $\gamma$ large would change the Daemon from an engine-like system into something more like a bomb.

The particle's motional energy $H_{\mathrm{M}}$ includes the usual kinetic energy with a mass $M$, plus a potential with dependence on position $q$ that can always be taken as linear over some range. The depot is modelled as an integrable subsystem with some high frequency $\Omega$, described in canonical action-angle variables $(\delta,I)$. The Daemon's particular case of (\ref{hamiltonian_daemon}) thus includes
\begin{align}
    H_{\mathrm{M}}(q,p) &= \frac{p^2}{2M} + Mgq, \\
    H_{\mathrm{I}}(\delta,I) &= \Omega I,
\end{align}
where $q$ is the particle's position, $p$ its momentum and the action variable $I$ of the depot is the ``fuel level'', while the angular variable $\delta$ is the phase that is canonically conjugate to the action variable $I$.

We take the potential as exactly linear in $q$ for simplicity; all our discussion can be adapted to slowly varying potentials adiabatically, and cases where the opposing force is frictional rather than conservative are not qualitatively different\cite{fieguth2022hamiltonian,fieguth2024dynamical}. For simplicity we also take the dependence of $H_{\mathrm{I}}$ on $I$ to be exactly linear, since our analysis likewise generalises adiabatically to (sufficiently weak) nonlinear dependence on $I$. 

The energy depot must have a finite capacity, and thus the fuel level is bounded by a minimum and maximum fuel level $\pm I_0$. These upper and lower bounds on $I$ must be respected by the coupling term $H_\mathrm{C}$, which will make $I$ time-dependent but can never make $|I|>I_0$.

\subsection{The nonlinear coupling}
\begin{figure}[t]
    \centering
    \includegraphics[width=0.47\textwidth]{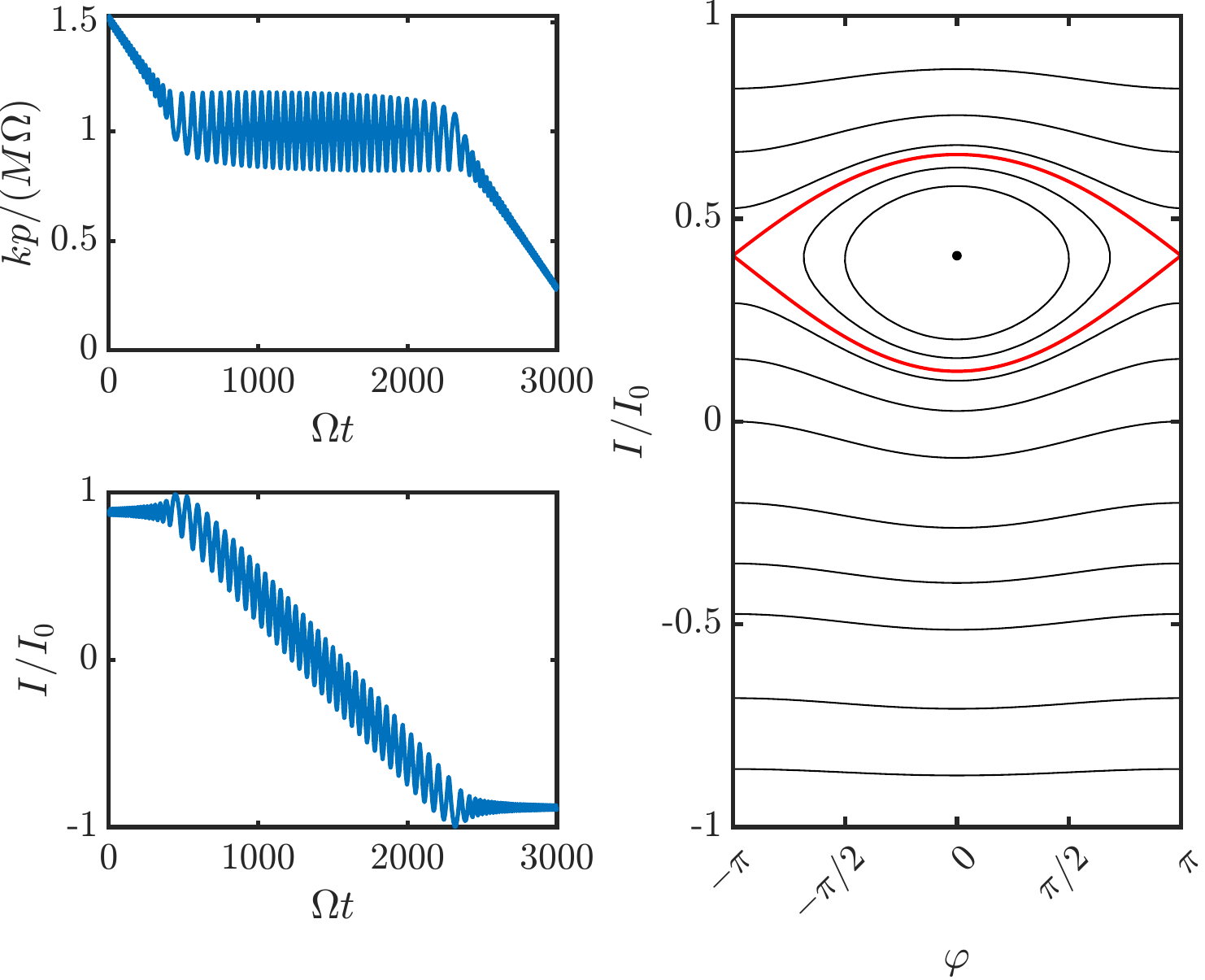}
    \caption{Left: Typical trajectory of the active particle's momentum $p$ and fuel level $I$ according to (\ref{hamiltonian_daemon}) over time during the active phase. The active phase is characterized by a steady energy flow from the particle's fuel tank degree of freedom $I$ to its momentum $p$ into work against an external opposing force. During this process, the fuel level decreases at a nearly constant rate, while the momentum is held at a nearly constant value. Right: Phase space of the transformed coordinate $(\varphi,I)$ with $\varphi\coloneqq kq- \delta$ with instantaneous contours at $\Omega t=1000$. The red contour shows the separatrix of the system in that instant, which is the boundary in phase space between the active and the inactive phases of the particle's dynamics. We chose here $Mg/(k\Omega I_0)=0.001$, $\gamma/\Omega=0.02$, $M\Omega/(k^2 I_0)=1$, $I(0)/I_0=0.9$, $kp(0)/(M\Omega)=1.5$.}
    \label{fig:Hamiltonian_daemon_examples}
\end{figure}

This requirement motivates the particular form of nonlinear coupling between particle's position $q$ and its depot's angle variable $\delta$ from \cite{PhysRevE.94.042127}, which  can be obtained by transforming a system of interacting particles:
\begin{align}
    H_{\mathrm{C}}(q,p,\delta,I) = \sqrt{I_0^2-I^2} \cos(kq-\delta),
\end{align}
for some constant $k$. This coupling provides a \textit{Chirikov resonance}\cite{Chirikov1971ResearchCT} which supports steady energy transfer between $H_{\mathrm{I}}$ and $H_{\mathrm{M}}$, but only in the narrow region of phase space where the argument of the cosine is approximately constant in time, i.e., $k\dot{q}-\dot{\delta}=0$. This condition defines the critical velocity $v_{\mathrm{c}}=\Omega/k$ at which the active particle can steadily move, against the opposing force $-Mg$, by steadily drawing energy from its depot. Figure \ref{fig:Hamiltonian_daemon_examples} shows a typical trajectory of the active particle's momentum $p$ and fuel level $I$ during the active phase (as well as slightly before and after it).

Further details of the Daemon system's behavior are supplied in \cite{PhysRevE.94.042127,fieguth2022hamiltonian,fieguth2024dynamical}. For our current problem, the most essential feature of the Daemon's active motion is that it eventually stops.

\subsection{End of active motion}

Since the internal depot energy is finite, active motion powered by the depot cannot continue forever. The secular rate at which $I$ falls, during active motion, is exactly the rate $\dot{I} = - Mg/k$ at which the loss of internal depot energy $\dot{H}_I = -Mg\Omega/k$ provides the power $(p_c/M)(Mg)=+Mg\Omega/k$ to keep moving at speed $p_c/M$ against the external force $-Mg$. The longest time for which the particle can continue to move actively is thus $t_m = 2kI_0/(Mg)$.

In \cite{PhysRevE.94.042127} it is shown that a certain adiabatic invariant generally requires the active phase to end before $I$ falls all the way to $-I_0$; it ends instead at some threshold $I\to I_*$ satisfying $-I_0< I_*<0$. This negative threshold $I_*$ is determined by the initial value of $I$ (in fact $I_* \doteq -I_i$ in the absence of viscous drag on the particle's motion). In \cite{fieguth2022hamiltonian} it is found that adding viscous drag against the active particle's motion actually lets the active phase persist longer, possibly until the depot is fully drained, by lowering $I_*$. With or without drag, however, it is easy to confirm by integrating the equations of motion numerically that the only thing which brings the active phase to an end is $I$ falling to some finite negative threshold. 

This suggests that if $I$ can simply be held above this negative threshold, by somehow restocking the energy depot as the particle moves, then the active phase could continue forever. It also seems intuitive that if the active particle interacts with crumbs that have more energy than it does, then it must on average be the winner in energy exchange with the crumbs, and thus be able to keep its depot energy above the threshold for continued active motion. Since this intuitive picture allows a Perpetual Motion Machine of the Second Kind, however, something must go wrong with it microscopically. To see what can go wrong, we introduce a mechanical model for crumbs that can exchange energy with the Daemon's depot.

\section{Microscopic harvesting model: feeding from crumbs}
\label{sec:hungry_daemon}
\subsection{The Hungry Daemon Hamiltonian}
We consider this total Hamiltonian with ``crumbs'' with which the ``Hungry Daemon'' active particle can interact:
\begin{align}
    H &= H_{\mathrm D} + \sum_n H_{n} \label{untransformierte_H} \\
    H_n &= \Omega J_n\\
    +& \kappa\, \theta(a/2-|q_n-q|) \sqrt{(J_0^2-J_n^2)(I_0^2-I^2)}\cos(\delta-\beta_n) \;,\nonumber
\end{align}
where $H_\mathrm{D}$ is still the Daemon Hamiltonian of (\ref{hamiltonian_daemon}), and each new canonical pair of action-angle variables $(\beta_n,J_n)$ represents a crumb at fixed location $q_n$, with $|J_n|\leq J_0$ for some $J_0$, in the same way that $|I|\leq I_0$. Each crumb $n$ is a small version of the active particle's internal depot; we will let all the crumbs have the same frequency $\Omega$ as the Daemon's internal depot, because this resonant case seems likely to be optimal for energy harvesting. The step function $\theta(a/2-|q-q_n|)$ means that the active particle only interacts with each crumb if it is within a distance $a/2$ from that crumb---and for simplicity we will ensure that no crumbs are within a distance $a$ from each other. The strength of the coupling between the active particle and the crumbs is given by $\kappa$.

The particular form of the coupling term in $H_n$ is chosen because it is equivalent to a classical spin-spin coupling $\propto (L_x L_{nx}+L_y L_{ny})$, in the minimal canonical representation of angular momentum:\begin{widetext}
\begin{align}
    \mathbf{L}&= \left(\begin{array}{c} \sqrt{I_0^2-I^2}\cos(\delta)
    \\ \sqrt{I_0^2-I^2}\sin(\delta) \\ I \end{array}\right),\qquad\qquad%\nonumber\\
    \mathbf{L}_n= \left(\begin{array}{c} \sqrt{J_0^2-J_n^2}\cos(\beta_n)
    \\ \sqrt{J_0^2-J_n^2}\sin(\beta_n) \\ J_n \end{array}\right)\;.
\end{align}
Although our numerical results in this paper will all now be based entirely on this specific form of coupling, as well as on the simplifying assumptions of equal parameters $J_0$, $\Omega$, $\kappa$ and $a$ for all crumbs, we will argue in our next Section that our conclusions are generic.

Some representative trajectories of this system for different \emph{non-thermal} choices of crumb parameters are shown in Figure \ref{fig:Hungry_daemon_examples}. The most important feature to note is that the Hungry Daemon can indeed successfully harvest energy from crumbs which all have high initial energies ($J_n(0)$ close to $+J_{0}$), but if it interacts instead with crumbs that all have low initial energies ($J_n(0)$ close to $-J_0$), then it actually loses energy to the crumbs, stalls sooner than it would if there were no crumbs, and does not replenish its internal depot.

The particular numerical results shown in Figure \ref{fig:Hungry_daemon_examples} do not immediately contradict the intuitive expectation that the active particle will always be able to harvest energy from crumbs if they have enough energy. Indeed, they may seem to support that intuition. However, it is precisely the pattern of high-energy crumbs yielding energy, while low-energy crumbs absorb it, which disproves the intuitive expectation and supports thermodynamics instead.

\begin{figure*}[t]
    \begin{subfigure}[b]{0.45\textwidth}
         \centering
         \includegraphics[width=\textwidth,trim={0 {1\textwidth} 0 {1\textwidth}},clip]{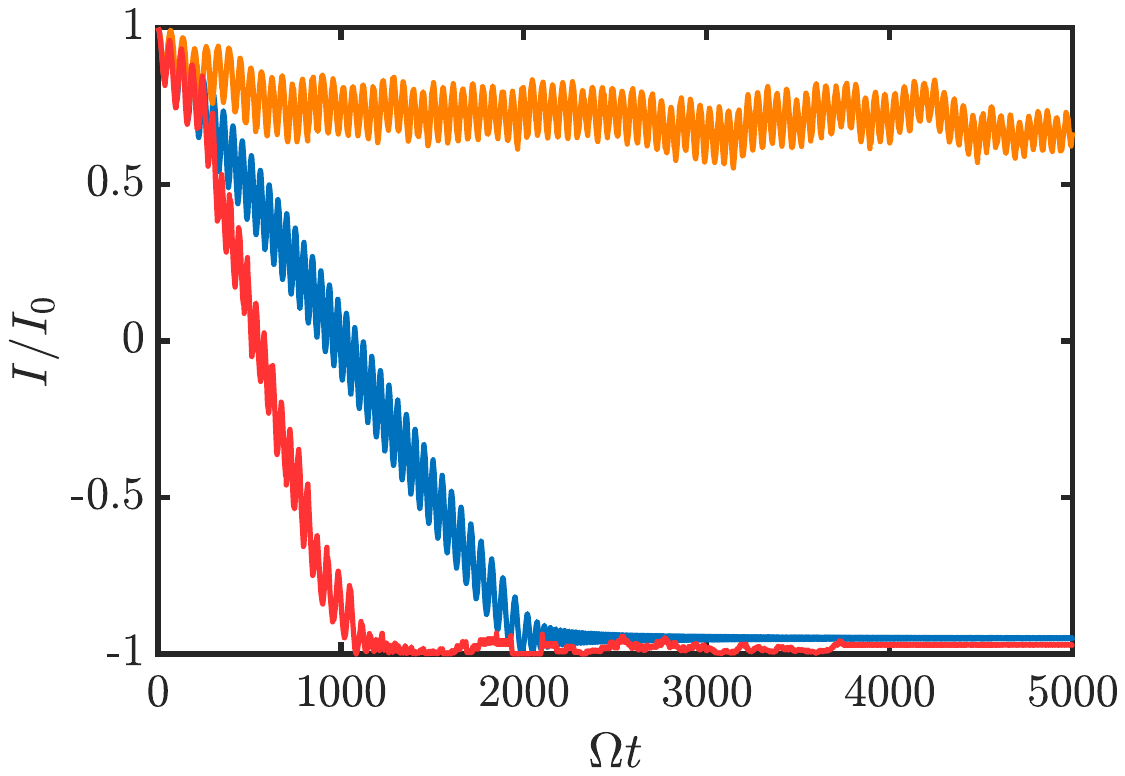}
     \end{subfigure}
     \begin{subfigure}[b]{0.45\textwidth}
         \centering
         \includegraphics[width=\textwidth,trim={0 {1\textwidth} 0 {1\textwidth}},clip]{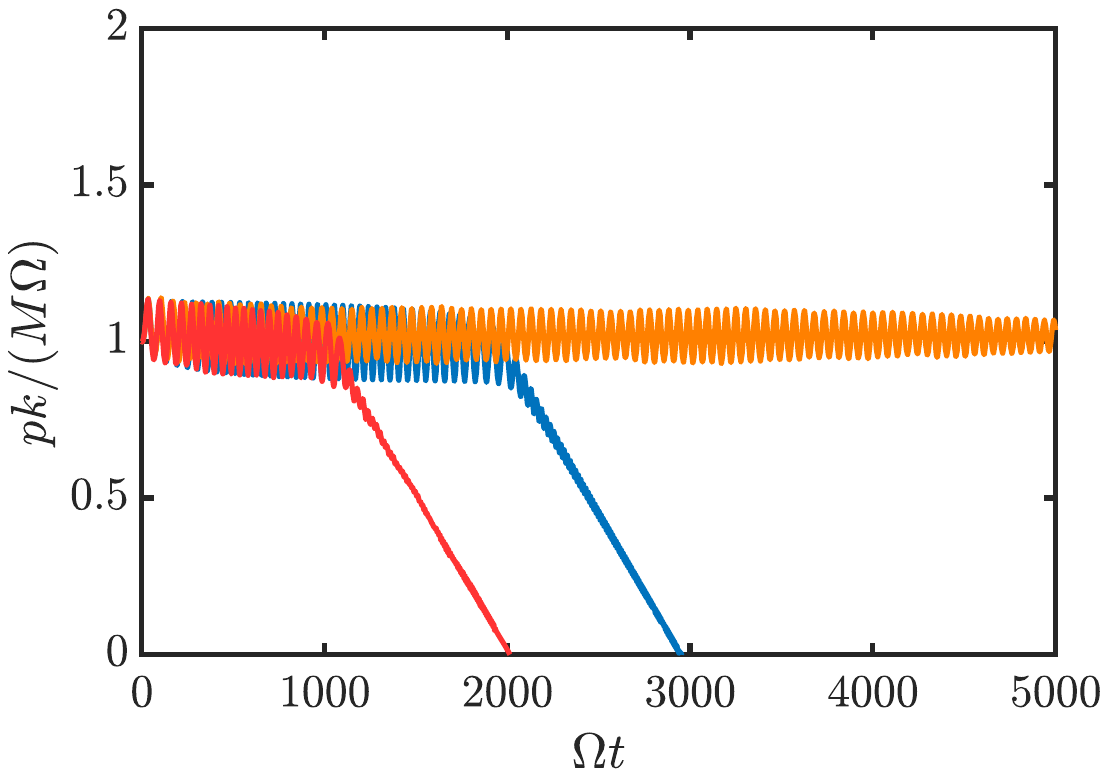}
     \end{subfigure}
    \caption{Typical trajectories of the active particle's internal depot ``fuel level'' $I/I_0$ and momentum $pk/(M\Omega)$, evolved over time according to (\ref{untransformierte_H}) in an environment without crumbs (blue), and with two different kinds of crumbs. The red and orange cases of crumbs are, respectively: small, low-energy crumbs having $J_n(0) = -0.6J_{0} \gg -I_0$ (red) and small, high-energy crumbs having $J_n(0) = +0.9J_0 \ll I_0$ (orange). The red path (with the low-energy crumbs) exits the active phase first, even sooner than in the case with no crumbs, and shows no sign of fuel replenishment. Meanwhile, the small, high-energy crumbs of the orange path sustain the active phase for at least the computed time span. All trajectories have $M=I_0k^2/\Omega$, $\gamma=0.02\Omega$, $g=10^{-3}\Omega^2/k$ and $a=1/k$. The crumbs are equidistantly spaced with $q_{n+1}-q_n=10/k$.}
    \label{fig:Hungry_daemon_examples}
\end{figure*}\end{widetext}

\subsection{Interaction with a single crumb in a simple limit}
\label{subsec:interaction_with_single_crumb}

We can obtain general analytical results for energy harvesting in our Hungry Daemon model by assuming an especially simple limit. This limit will then show us what \emph{can} go wrong with the intuitive argument in favor of perpetual motion from thermal crumbs, such that instead thermodynamics is right. After we understand this issue in the simple limit, we will then argue in the following Section that the problem with the intuitive argument is not restricted to the simple limit, but is instead a general implication of Liouville's Theorem.

We consider this simple limit by imposing the following special conditions on our Hungry Daemon model parameters:
\begin{itemize}
    \item The crumbs are much smaller energetically than the particle's internal depot: $J_{0} \ll I_0$, as for the red and orange curves in Figure \ref{fig:Hungry_daemon_examples}, and as one expects for individual food particles consumed by a small mobile organism;
    \item The typical interaction duration $\Delta t = a/v_c$ between the actively moving particle and a crumb is short compared to $1/\gamma$, so that $H_\mathrm{C}$ can be neglected over this short interval; and
    \item The effective crumb-depot coupling strength $\sim\kappa J_0$ is, on the other hand, large enough that the interaction between a crumb and the depot is \emph{not} negligible while $|q-q_n|\leq a/2$.  
\end{itemize}
In this simple limit the Hungry Daemon active particle effectively moves just as if there were no crumbs, except that its depot energy $\Omega I(t)$ can make a small jump whenever it passes one of the $q_n$ crumb locations.

This jump in depot energy occurs over the short interaction time with a crumb, while $q(t)$ is within the interaction range $a/2$ of a crumb location $q_n$. In that time, the crumb reservoir and the particle depot effectively evolve under the simple coupling Hamiltonian 
\begin{equation}
    H\to H_\mathrm{int} = \kappa\sqrt{I_0^2-I^2}\sqrt{J_{0}^2-J_n^2}\cos(\delta-\beta_n)\;. \label{interaction_Hamiltonian}
\end{equation}
This effective Hamiltonian $H_\mathrm{int}$ exactly conserves $I+J_n$, so that the jump in $I(t)$ over the short interval $\Delta t$ is exactly the negative of the change in $J_n(t)$ over this interval. Since $|J_n|\leq J_0\ll I_0$, furthermore, we can neglect the proportionally small possible change in $I(t)$ and treat $I(t)$ as constant throughout this short interval. With this approximation it becomes easy to solve the canonical equations of motion for $J_n(t)$. Defining $t_n$ as the moment when the active particle begins interacting with crumb $n$, we obtain
\begin{widetext}
\begin{align}
    J_n(t) &=  J_n(t_n)\cos\Big(\kappa\sqrt{I_0^2-I^2(t_n)}(t-t_n)\Big) %\nonumber\\
    + \sqrt{J_0^2-J_n^2(t_n)}\sin[\beta_n(t_n)-\delta(t_n) ]\sin\left(\kappa\sqrt{I_0^2-I^2(t_n)}(t-t_n)\right)\;.
\end{align}

We can safely assume that the relative phase $\beta_n(t_n)-\delta(t_n)$ is random, first of all because all $\beta_n$ have the same energy until the active particle actually starts interacting with crumb $n$, and secondly because $\delta(t)$ evolves rapidly while the active particle moves between crumbs. On average over an ensemble of many crumbs which all have arbitrary $\beta_n-\delta$, therefore, the jump in the active particle's depot level $I(t)$ after interacting with a crumb with initial energy $J_n(t_n)$ for a duration $\Delta t$ will be
\begin{align}\label{DIn}
    &\Delta I_n \coloneqq \langle I(t_n+\Delta t)-I(t_n)\rangle = \langle J_n(t_n)-J_n(t_n+\Delta t)\rangle\nonumber\\
    =& J_n(t_n)\left[1-\cos\left(\kappa\sqrt{I_0^2-I^2(t_n)}\Delta t\right)\right] \doteq J_n(t_n)[I_0^2-I^2(t_n)]\frac{(\kappa\Delta t)^2}{2}\;,
\end{align}\end{widetext}
where in the last step we have used the assumption that $\kappa I_0 \Delta t$ is small.

The crucial feature of $\Delta I_n$ according to (\ref{DIn}) is that it is directly proportional to $J_n(t_n)$. Since the Daemon active particle continues to move in the same direction during its active motion, the active particle will not have interacted with any given crumb until time $t_n$, and since $J_n(t)$ is conserved until the active particle first comes within $a/2$ of its location $q_n$, $J_n(t_n)$ is simply the initial value of $J_n$. At least in this simple limit of our particular Hungry Daemon model, therefore, whether the active particle will on average gain energy from a crumb, or lose energy to the crumb, does \emph{not} depend on whether the crumb is initially more energetically excited than the active particle's depot, but only on whether the crumb is closer in energy to its own energy maximum or to its own energy minimum.

\subsection{Second Law of Thermodynamics}
Over a short interaction time $\Delta t$ we may assume that the actively moving Daemon keeps moving at close to its active speed $\Omega/k$, so that the interaction time can be computed self-consistently as
\begin{equation}
    \Delta t = \frac{ka}{\Omega}\;.
\end{equation}
Inserting this in (\ref{DIn}) and averaging over a thermal distribution of $J_n(t_n)$ at temperature $T$, we find the average energy gain from interacting with crumb $n$ during active motion to be
\begin{align}
   {\Delta E}(t_n) &= \int_{-J_0}^{J_0}\!dJ_n\,\frac{e^{-\frac{\Omega J_n}{k_B T}}}{2\frac{k_B T}{\Omega}\sinh\frac{\Omega J_0}{k_B T}}\,\Omega\Delta I_n\nonumber\\
    &= \frac{(\kappa k a)^2J_0}{2\Omega}\left(\frac{k_B T}{\Omega J_0}-\coth\frac{\Omega J_0}{k_B T}\right)\Big(I_0^2-I^2(t_n)\Big)\;.
\end{align}

We can immediately note that $\Delta E(t_n)$ is \emph{negative} for all finite positive $T$, because $\coth x > 1/x$ for all $x>0$. On average the Hungry Daemon does \emph{not} succeed in harvesting energy from thermally excited crumbs at any positive temperature. Instead, its internal depot actually loses energy to the crumbs, on average. At least within the simple limit and its approximations, therefore, mechanics agrees with thermodynamics that perpetual motion cannot be powered solely by a thermal source with a single positive temperature.

\begin{figure}
    \centering
    \includegraphics[width=0.5\textwidth]{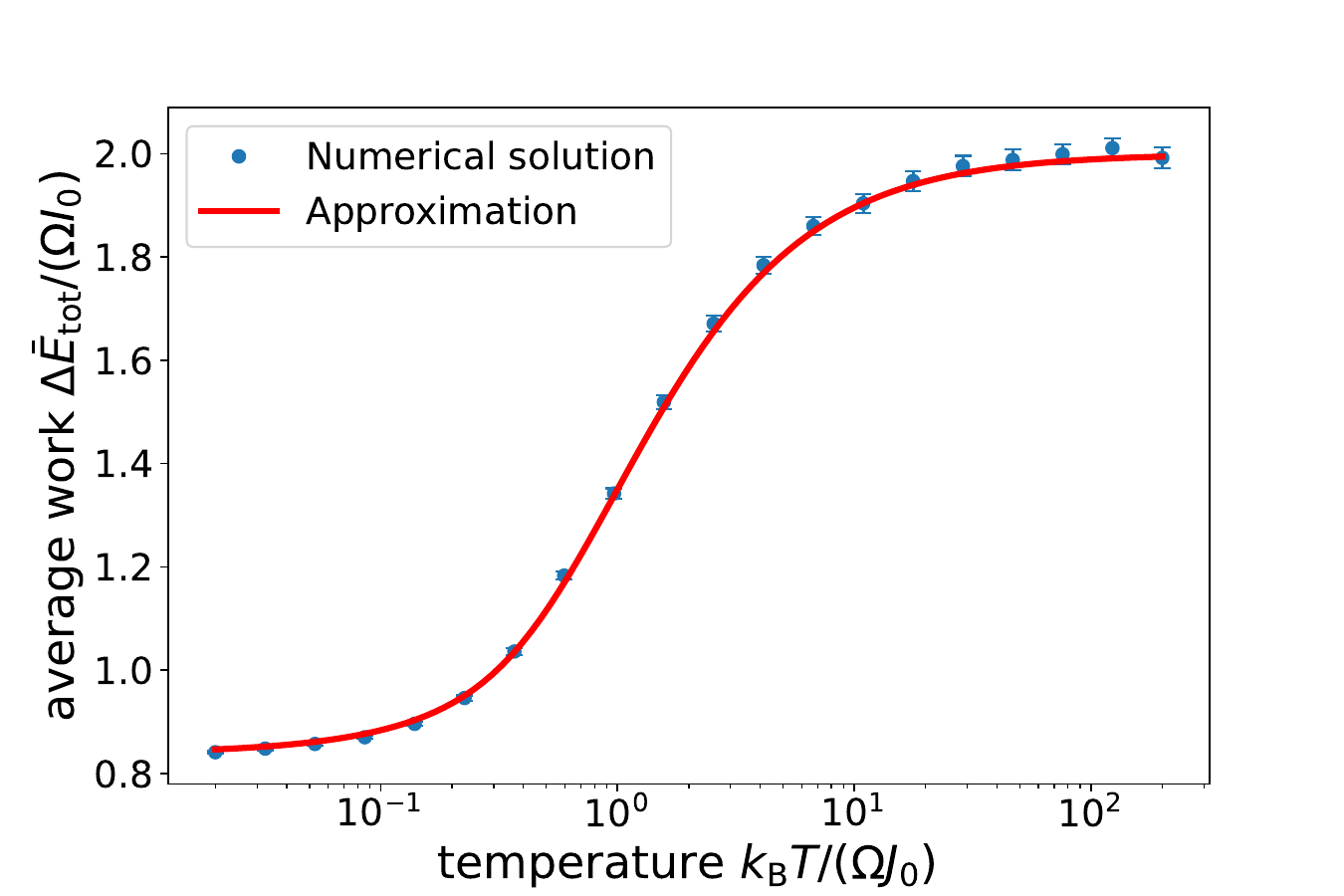}
    \caption{Average total work $\Delta\bar E_{\text{tot}}$ of the Daemon approximated according to (\ref{Idot}) (red curve) together with the results of a Monte Carlo computation with 1000 trials (blue points with 95\,\% confidence intervals as error bars) in dependence on temperature of the crumb energy distribution. The close agreement is unsurprising for the specific case shown, which has $M\Omega/(k^2 I_0)=1$, $\gamma=0.02\Omega$, $g=10^{-3}\Omega^2/k$, $J_{0}=0.05I_{0}$, $a=1/k$, $\Delta q = 10/k$, $\kappa =\Omega/I_0$ $I(0)=I_0$ and $p(0)=M\Omega/k$.}
    \label{fig:delta_E_T_plot}
\end{figure}

We can confirm that our approximations which yielded this result are indeed accurate, by further considering that the Daemon moves actively through an ensemble of crumbs spaced $\Delta q$ apart, and therefore interacts with crumbs at an average rate $\Delta q k/\Omega$. Including the effect of the opposing force $-Mg$, the average total rate at which the Daemon's internal depot must be changing should then be
\begin{equation}\label{Idot}
    \dot{I}(t) = -\frac{Mg}{k}+\frac{\Delta q k}{\Omega}\frac{{\Delta E}(t)}{\Omega}\;,
\end{equation}
which is a Ricatti equation for $I(t)$. If we start the Daemon with a full internal depot, $I(0)=I_0$, then it runs actively until the time $t^*$ at which $I(t^*)=-I_0$, and we can find an analytical expression for $t^*$ by integrating (\ref{Idot})\;. Multiplying by the power $Mg\Omega/k$ applied by the internal depot against the external force yields the average total work $\Delta\bar{E}_\text{tot}$ done by the Daemon in its active motion; this analytical result for the Daemon's work output, based on our approximate analysis in this section, is compared in Fig.~\ref{fig:delta_E_T_plot} with a numerically exact solution to the full equations of motion for our system, for ensembles of random crumbs obtained by sampling thermal distributions at different temperatures. The agreement is close.

\section{Discussion}

\subsection{Generality?}
Our analytical and numerical results have only explicitly described one simple limit of one artificial model system. By showing what can go wrong with  intuitive expectation, however, these model calculations have indicated the general mechanical reason why the Second Law must be valid for a large class of possible energy-harvesting active particles. The mechanical reason for the Second Law, at least in this class of systems, is Liouville's theorem.

For crumbs having any possible Hamiltonian $H_n$, any positive-temperature phase space distribution 
\begin{equation}
	P_n(\vec{Q}_n,\vec{P}_n) = Z_n^{-1} \exp\Big(-\frac{H_n(\vec{Q}_n,\vec{P}_n)}{k_B T}\Big)
\end{equation}
can be considered as an ensemble of sub-ensembles that each have uniform probability for all energies below some maximum energy $\mathcal{E}$:
\begin{equation}
	P(\vec{Q}_n,\vec{P}_n) = \int\!d\mathcal{E}\tilde{P}_n(\mathcal{E})\theta(\mathcal{E}-H_n)
\end{equation}
for some sub-ensemble probability distribution $\tilde{P}_n(\mathcal{E})$.

For each such sub-ensemble of all crumbs, each uniformly filling the phase space of each crumb $n$ from the ground state to some $\mathcal{E}_n$, and for any initial state of the active particle and its internal depot, the entire system of crumbs and active particle fills a certain total phase space volume uniformly. This will be true even for greatly generalized versions of our simple model, with arbitrary forms of interaction between active particle and crumbs. 

Liouville's theorem then says that this initial phase space volume (actually the product of all the individual subsystem volumes) can never change under time evolution. Yet if the active particle were to consistently extract energy from the crumbs, on average, their ensembles would have to be compressed downwards from $\mathcal{E}_n$ to their ground states. If the energy harvesting could continue indefinitely, then this shrinking of the initial phase space volume would have to continue indefinitely, as more and more crumbs and their energies drained---and therefore their phase space volumes.

The internal depot of an active particle is finite, and can only hold a finite amount of energy---or phase space volume. The momentum of the active particle likewise remains bounded, and its position is supposed to be systematically moving in time, if perpetual energy harvesting is happening, not diffusing over an exponentially increasing range. So there is no way for the phase space volume of the active particle itself to increase sufficiently to make the steady decrease of crumb phase space volume consistent with Liouville's theorem.

Phase space volume and Liouville's theorem thus function for active particle energy harvesting as the microscopic, mechanical proxies for entropy and the Second Law. This is the relationship which is always suspected, in seeking the microscopic underpinnings of the Second Law, but it cannot always be confirmed so explicitly. The impossibility of perpetual energy harvesting from thermally excited crumbs, which we found in our particular model, is indeed clearly general. As a new variant of a Maxwell's Demon, the Hungry Daemon has indeed shed some light on the relationship between thermodynamics and microscopic mechanics.

\subsection{\texorpdfstring{Harvesting can succeed for $T<0$}{Harvesting can succeed for T<0}}

The Hungry Daemon also confirms microscopically and mechanically something that is generally assumed about energy harvesting: that it does \emph{not}
actually violate thermodynamics, because energy that is available to be harvested from an environment is \emph{not} simply heat. For crumbs with energy bounded from above as well as from below, negative temperatures are possible. For negative temperatures, $\Delta I_n$ in Eqn.~(\ref{DIn}) can be positive, and for given opposing force $-Mg$ and rate of encountering crumbs there can be a sufficiently large negative $T$ for the crumbs at which the Hungry Daemon does replenish its depot fast enough to keep moving indefinitely. And with negative temperature, the Second Law does \emph{not}
forbid perpetual motion, since extracting energy then \emph{increases} entropy.

Crumbs with Boltzmann-distributed energies at a negative $T$ are not necessarily realistic, even if the crumb energy is bounded from above, but the formal possibility of perpetual motion through energy harvesting from negative-temperature crumbs indicates that energetic population inversion of any kind can be sufficient to permit energy harvesting. With negative temperature or with any more general kind of inversion, the sub-ensemble argument that we have just presented must be revised to consider sub-ensembles of uniform probability \emph{above} a threshold energy $\mathcal{E}$, instead of below. The argument from Liouville's theorem then changes direction, to forbid steady energy loss to the crumbs, but allow steady harvesting. Liouville's theorem does not allow uniformly filled phase space volumes to increase, any more than it allows them to decrease, but it allows an initially uniform ensemble to become dilute, being made porous with swirls, so that it fills a larger volume in a coarse-grained sense. It does not allow a uniformly filled phase space volume to compress. So energy harvesting \emph{is} allowed to continue indefinitely, if the crumbs have sufficient energetic inversion.

In this sense the Hungry Daemon model has also shed light on energy harvesting, by highlighting the special thermodynamical character of food. Food must not merely be energy, but thermodynamically available energy: it must have some form of population inversion, with higher crumb energies appearing more often than they would in any positive-temperature thermal distribution.

\subsection{Heat and work and intuition}
The intuitive expectation for the Hungry Daemon was that it would surely gain energy from crumbs, at least on average, if they had more energy or higher temperature than it did. In retrospect this was a self-inconsistent expectation. It assumed that energy exchange between a small organism and crumbs would be like heat exchange between macroscopic bodies, and yet it then implied a contradiction of the Second Law for heat exchange. 

This is perhaps a useful caution for phenomenological models like active particles. It can certainly be useful to make simple, concrete models for phenomenologically described systems, but we must be careful not to extend our intuitive expectations of phenomenology beyond their regime of validity, lest we build fundamental impossibilities into our models.

The whole point of Thermodynamics is that heat is a form of energy, but it is a special form of energy that obeys the special law of entropy increase. Heat can take many microscopic forms, but not all energies behave as heat. To make these subtleties clear, abstract and axiomatic definitions of heat and work can be useful in thermodynamics and statistical mechanics. Purely mechanical analysis, such as we have offered in this paper, can add complementary insights as well.

\subsection{Energy-harvesters versus heat engines}

In order to satisfy Liouville's theorem, energy extraction must conserve action. Re-examining our impossibility proof from Liouville's theorem, above, we can see that energy harvesting from a single-temperature source could become possible after all, if the active particle would just divert a bit of its harvested energy to steadily excite some additional, ancillary degrees of freedom in its environment which were previously less energetically excited than the crumbs. The amount of diverted energy that would be needed to satisfy Liouville's theorem, and enable steady energy harvesting, does not need to be as much as the energy which is taken from the crumbs. It just has to be enough to expand the phase space volume of those ancillary degrees of freedom enough to satisfy Liouville.

In other words, we can harvest energy from a single hot bath, if we can expel heat to a colder bath. We have invented the heat engine. Or rather, of course, we have recognized that having access to two baths at different temperatures, enabling a heat engine, is a particular form of the kind of uneven distribution of energies in the environment that can also be exploited for energy harvesting from food or fuel. An energy-harvesting active particle is thus qualitatively like a heat engine, up to a point, in that it requires a certain population inversion, though it is different in the specific form of inversion it needs. Heat is a special form of energy, and so are food and fuel: in particular, food and fuel are special in that they must \emph{not} be heat. They need to have some form of inverted distribution of energies, with higher energies more likely rather than less, at least in some range; they cannot just have a thermal distribution with a single positive temperature.

\section*{Acknowledgment}

S.B.~gratefully acknowledges J.~Bohm, C.~Mink, C.~Dittrich and M.~Will for valuable discussions as well as T.~Pfeifer and E.~Wamba for their constructive feedback on this paper. D.M.F.~and J.R.A.~acknowledge support from the Deutsche Forschungsgemeinschaft (DFG) through SFB/TR185 (OSCAR), Project No.~277625399.

\appendix
\begin{widetext}

\section*{Appendix}

\section{Equations of motion}

The resulting EOM from the Hamiltonian (\ref{untransformierte_H}) are given by
\begin{align}
    \dot{q} &= \frac{\partial H}{\partial p} = \frac{p}{M}, \label{q} \\
    \dot{p} &= -\frac{\partial H}{\partial q} = 
    -Mg - k\gamma\sqrt{I_0^2-I^2}\sin(kq-\delta), \label{p} \\
    \dot{\delta} &= \frac{\partial H}{\partial I} = \Omega + \gamma \frac{I}{\sqrt{I_0^2-I^2}}\cos(kq-\delta) -\kappa \sum\limits_{n=1}^N \theta(a/2 - |q-q_n|) I \sqrt{\frac{(J_0^2-J_n^2)}{(I_0^2-I^2)}}\cos(\delta-\beta_n), \\
    \dot{I} &= -\frac{\partial H}{\partial \delta} = \gamma\sqrt{I_0^2-I^2} \sin(kq-\delta) + \kappa\sum\limits_{n=1}^N \theta(a/2 - |q-q_n|) 
    \sqrt{(J_0^2-J_n^2)(I_0^2-I^2)}\sin(\delta-\beta_n), \label{I} \\
    \dot{\beta_n} &= \frac{\partial H}{\partial J_n} = \Omega -
    \kappa \theta(a/2 - |q-q_n|) J_n \sqrt{\frac{I_0^2-I^2}{J_0^2-J_n^2}}\cos(\delta-\beta_n), \\
    \dot{J_n} &= -\frac{\partial H}{\partial \beta_n} = -\kappa \theta(a/2 - |q-q_n|) \sqrt{(J_0^2-J_n^2)(I_0^2-I^2)} \sin(\delta-\beta_n). \label{J_n}
\end{align}
\end{widetext}

\end{document}